\documentclass{cjpsuppl}
\input epsf.tex
\begin{document}
\title{Experimental Status of Beyond the Standard Model Collider Searches}
\authori{M.\,Spiropulu}
\addressi{Physics Department,
CERN CH 1211, Gen\`{e}ve 23, Suisse}
\authorii{}   \addressii{}
\headtitle{Experimental Status of Beyond the Standard Model Searches
}
\headauthor{M. Spiropulu}
\lastevenhead{M. Spiropulu: Experimental Status  
of Beyond the Standard Model Searches}
\pacs{12.60.-i}
\keywords{Colliders, Exotics, Supersymmetry, Extra Dimensions}
\refnum{}
\daterec{20 November 2004;\\final version 17 January 2005}
\suppl{A}  \year{2004} \setcounter{page}{1}
\maketitle

\begin{abstract}
This is a brief review of experimental
strategies for physics beyond
the Standard Model based on the talk given in the ``Physics at LHC'' in Vienna, July 2004 \cite{vienna}. The emphasis is on Tevatron thematology and experience.

\end{abstract}

\section{The high energy collider data to-date}
The world-wide high energy collider data composition to-date is:
\begin{itemize}
\item proton-antiproton collisions
	\begin{itemize}
	\item $\sim 120$ pb$^{-1}$ at $\sqrt{s}= $ 1.8  TeV.
	\item $\sim 300$ pb$^{-1}$ at  $\sqrt{s}= $ 1.96  TeV.
	\end{itemize}
\item electron-proton collisions 
	\begin{itemize}
	\item $\sim 200$ pb$^{-1}$ at $\sqrt{s}= $ 320  GeV.
	\end{itemize}
\item electron-positron collisions
	\begin{itemize}
	\item $\sim 1$ fb$^{-1}$ at $\sqrt{s}$ up to 209  GeV.
	\end{itemize}
\end{itemize}

Results obtained by the experiments recording the data above
are well in agreement with 
the Standard Model predictions. Of what I call our ``high 
energy delinquencies'' today,  two I consider collider-wise curious. 
The first one is pending, and data from the Tevatron Run II 
will answer it: it is the need for a more precise top quark mass 
measurement from the CDF and D\O\ collaborations \footnote{The individual 
Run I most precise top mass measurent was reported by the D\O\ collaboration 
to be $m_{top}=180.1\pm 3.6 \pm 3.9 GeV/c^{2}$ \cite{nature}. The Run II most precise one
was recently reported by the CDF Collbarotion to be $m_{top}=173.5 + 4.1 - 4.0 GeV/c^{2}$
\cite{cdf-top}.}. 
The second is the
understanding of the $A_{fb}^{0b}$ measurement at LEP, covered in this workshop by 
Paul Langacker \cite{PL}. The former is critical in predicting new physics and 
the latter may suggest new physics affecting the third family.

Other excesses in different channels observed at CDF Run I/Run II, LEP, and
HERA have been assessed statistically not significant. The theoretical
and phenomenological part on beyond the standard model 
searches in the same workshop (by Joe Lykken, see these proceedings \cite{JL-vienna}),  
covers in detail the exotic tendencies today in collider
physics. I would like to re-emphasize that most all are 
conceived within one kind or another of extra 
dimensions \cite{JS} and supersymmetric scenarios \cite{SUSY}.  The
strict or loose dualities between different frameworks
for physics ``beyond the standard model''
have a direct experimental consequence:
the final states and signatures of the models are very
similar. This renders the characterization of an excess or a deviation,
a fine and probably long challenge. To mention a couple of examples:
the question ``is it universal extra dimensions \cite{UED} or is it SUSY?''
or ``is it a Randall-Sundrum \cite{lisa} graviton mode or a Z$^{\prime}$ \cite{JR}''
is not going to be answered immediately when the excess is observed.
The results from all the collider data we have, together with
the as yet unobserved Higgs, and in concilience
with the data on the neutrino masses and the composition
of the universe, point to a remainder in particle
physics. But they do not point to the nature of it. There is something
(probably a lot) more but it is tricky to say what it is. 
In high energy physics
today when we talk about beyond the standard model 
phenomena, including supersymmetry, we talk about data 
at the edges or tails of the standard model distributions, be it
large invariant masses or tails of missing transverse energy.
It has become a clich\'{e} (albeit wise) that the accurate and precise determination
of the standard model physics is crucial as a background to direct exotic 
searches and as an indirect probe of new physics. 

\section{The signature-model correspondence}

The plethora of what the CDF collaboration (used as an example
of the Tevatron experiments) calls ``very exotic'' searches is presented in the 
indicative listing (circa spring 2004) of signatures (corresponding to models) explored below:

\begin{itemize}
\item Di-Lepton Resonances
  \begin{itemize}
  \item using $ee,~\mu\mu,~ \tau\tau$
  \item searching for $Z^{\prime}$, RS Extra Dimensions, Technicolor
  \end{itemize}
\item Same-Sign Di-Lepton Resonances
  \begin{itemize}
 \item using $ee,\mu\mu, e\mu, \tau\tau$  
 \item searching for $H^{++}$
  \end{itemize}
\item Di-Lepton$+$Photon
  \begin{itemize}
  \item using $ee\gamma,\mu\mu\gamma, \tau\tau\gamma$ 
  \item searching for heavy leptons
  \end{itemize}
\item Di-Lepton$+$Di-jet
  \begin{itemize}
  \item using $eejj$,$\mu\mu jj$, $\tau\tau~ jj$, $e\nu jj$, $\mu\nu jj$, $\tau\nu jj$, $\nu\nu jj$ 
  \item searching for leptoquarks
  \end{itemize}
\item Photon+missing $E_{T}$
  \begin{itemize}
  \item using $\gamma+ \rm{missing}~E_{T}$ 
  \item searching for ADD (see \cite{JL-vienna} and \cite{ADD}) Graviton
  \end{itemize}
\item Photon+jet
  \begin{itemize}
  \item using $\gamma+$ $b$-jet 
  \item searching for $b^{\prime}$
  \end{itemize}
\item Highly-ionizing (slow) track
  \begin{itemize}
  \item searching for $H^{++},H^{--}$, monopoles, UEDs, 
   stops and staus  (and charged split SUSY-type R-hadrons 
   more recently \cite{split-savas}, \cite{split-savas1}).
  \end{itemize}
\end{itemize}
	
More signatures has been added to the list since, and are being investigated with the data. 

Taking the reverse  route, a particularly fashionable example is the
signatures generated within the ADD model. 

\begin{table}
\medskip
\begin{center}
\begin{tabular}{lll}
\hline\hline
\multicolumn{3}{c}{direct graviton production}\\
\hline\hline
$e^{+}e^{-}, p \bar p$ & $\longrightarrow$      & $\gamma G,~ jG$    \\
			&		    & $ZG$\\
			&		    & $WG$\\
			&		    & $f\bar f G$\\
\hline\hline
\multicolumn{3}{c}{virtual graviton exchange}\\
\hline\hline
$e^{+}e^{-}, p \bar p$&  $\longrightarrow$ & $\gamma \gamma,~WW,~ZZ$\\ 
$q\bar q $ & & $\ell^{+}\ell^{-}$\\
$\ell^{+}\ell^{-}$&  & $q\bar q$\\
$ep$ &  & $eX, ~\nu X$\\
$q\bar q$ &  & $jj, ~t\bar{t}$ \\
\hline\hline
\end{tabular}
\caption{\label{table3} Signatures generated within the ADD 
model in the cases of direct graviton emission and graviton
exchange. $G$ denotes a Kaluza-Klein graviton.\hbox to100pt{}} 
\end{center}
\end{table}

It is notable that LEP high energy experimentalists 
produced results on these searches almost as soon as the scenarios appeared:
Higgs 
( e.g. visible mass analyses $e^{+}e^{-}\longrightarrow Z+$ missing energy)
and GMSB type analyses (e.g  $e^{+}e^{-}\longrightarrow \gamma+$ missing energy)
were turned practically overnight into searches for direct $G$ production in the ADD model.
Anomalous $Z\gamma\gamma$ couplings, $WW$, $Z\gamma$ analyses  
(e.g. $e^{+}e^{-}\longrightarrow \gamma\gamma,~VV$),
were applied in  searches for virtual $G$ exchange  effects and so did
analyses with Bhabhas and other QED type of measurements.

The case of asymmetric or TeV$^{-1}$ extra dimensions 
(\cite{ignatios} also see J. Lykken's review in these proceedings \cite{JL-vienna}
and references therein) offers
similar signatures. In this case Kaluza-Klein $Z$, photon or gluon  exchange
affects the di-lepton, di-photon or di-jet cross sections at
high $p_T$. Drell-Yan production at the Tevatron, HERA NC and CC deep inelastic scattering
analyses, hadronic and leptonic cross sections and angular distributions at
LEP 2, have all been studied by Cheung and Landsberg 
in the context of TeV$^{-1}$ extra dimensions \cite{tev-5}. 
The limits obtained are shown in table \ref{greg}.  
The overall limit on the compactification scale, $M_C > 6.8$ TeV has improved the one from the electroweak precision data.
The estimated sensitivity reach at the Run II of the Tevatron
and at the LHC   using the Drell-Yan process  is 2.9 TeV with 2 fb$^{-1}$
of $p \bar{p}$ collisions at $\sqrt{s}=1.96$ TeV and 13.5 TeV with 100 fb$^{-1}$
of $pp$ collisions at $\sqrt{s}= 14$ TeV  (and assuming 3\% overall 
uncertainty from systematics)  correspondingly. 
Bal\'{a}zs and Laforge \cite{balazs} showed that using the di-jet 
production, the LHC can probe $M_C \sim 5-10$ TeV. A Run II 
search at D\O\ using the invariant mass of di-electrons  from 200 pb$^{-1}$
(shown in figure \ref{d0-diem}) yields a 95\% CL lower limit on 
$M_C$  of 1.12 TeV.

\begin{table}[htb]
\medskip
\begin{center}
\begin{tabular}{lc}
      &   $M_c^{95}$ (TeV) \\
\hline \hline
LEP~2:                       & \\
{} hadronic cross section, ang. dist., $R_{b,c}$
    & 5.3 \\
{} $\mu,\tau$ cross section \& ang. dist.
     & 2.8 \\
{} $ee$ cross section \& ang. dist.
      & 4.5\\
{} combined      &
6.6 \\
\hline
HERA:     &  \\
{} NC     &   1.4 \\
{} CC      & 1.2 \\
{} HERA combined  &1.6 \\
\hline
TEVATRON I (120 pb$^{-1}$) :               &  \\
{} Drell-yan           &  1.3 \\
{} Tevatron di-jet      &   1.8 \\
{} Tevatron top production &   0.6 \\
{} Tevatron combined     &2.3 \\
\hline \hline
All combined & 6.8 \\
\hline
\end{tabular}
\caption{\label{greg}  95\% CL upper limits on $M_c$ for
individual data sets and combinations.} 
\end{center}
\end{table}

\section{Di-objects}

In the example of searches using di-leptons in the final state
resulting from an exotic object produced in $p\bar{p}$ 
collisions we note that the signature is usually well defined and triggered: 
two energetic, isolated, same flavor,
opposite sign leptons. The summary of the Tevatron (CDF specific in this case)
experience for this class of searches is that the Drell-Yan background, 
although  irreducible, is well simulable and calculable and estimated to 5\%.
The remaining uncertainty is
mainly from resolution and acceptance since, after normalizing to the
$Z$, the luminosity uncertainty drops out. At high invariant mass 
the dominant background and background uncertainty component 
is jets misidentified as  electorns. 
Other ``Fake'' lepton backgrounds, {\it i.e.} pions decaying in flight, 
conversions, $K^{+}\rightarrow\mu\nu$ as well as 
heavy flavor ($b\rightarrow c\ell\nu$) are  not predicted but 
estimated from control data samples to $\sim$30-50\%.  
Cosmics in the muon channels have been always more of a 
problem than one might think for the Tevatron experiments and
 are estimated only to $\sim$30-50\%. $W$+jets, di-bosons and top 
backgrounds are eliminated with  a high invariant mass requirement.
In general the di-object exotic searches look for a resonance 
or a deviation in the di-object invariant mass spectrum, a cross section
excess at large $p_T$, and modifications
in the angular distribution of the final state objects especially 
at high invariant masses.

\begin{figure}
\centerline{\epsfxsize 3.55 truein \epsfbox {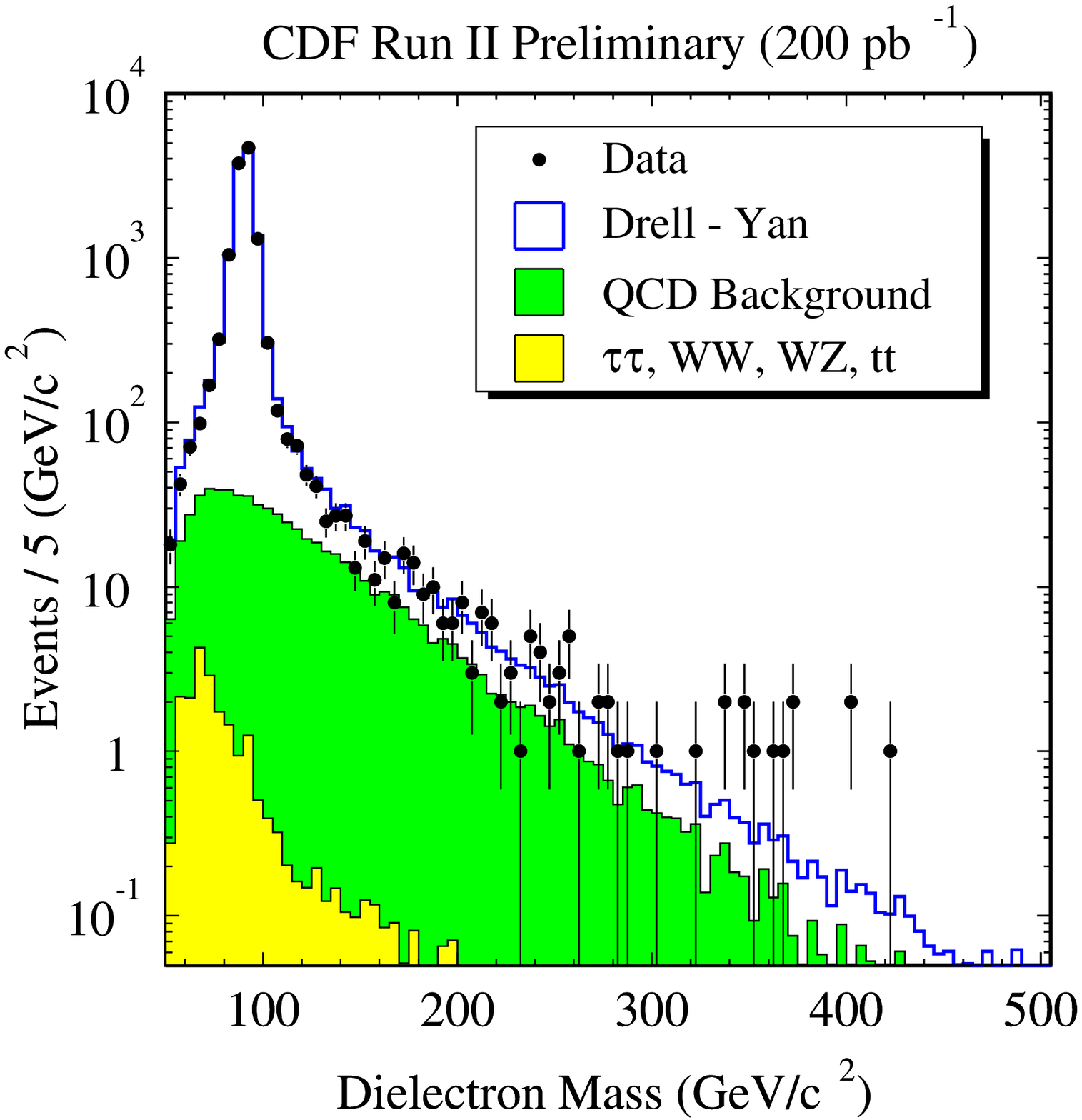}}
\caption{\label{cdf-ee} Di-electron invariant mass spectrum comparison
between data and estimated backgrounds at CDF RunII.
\hbox to244pt{}}
\end{figure}

\begin{figure}
\centerline{\epsfxsize 3.55 truein \epsfbox {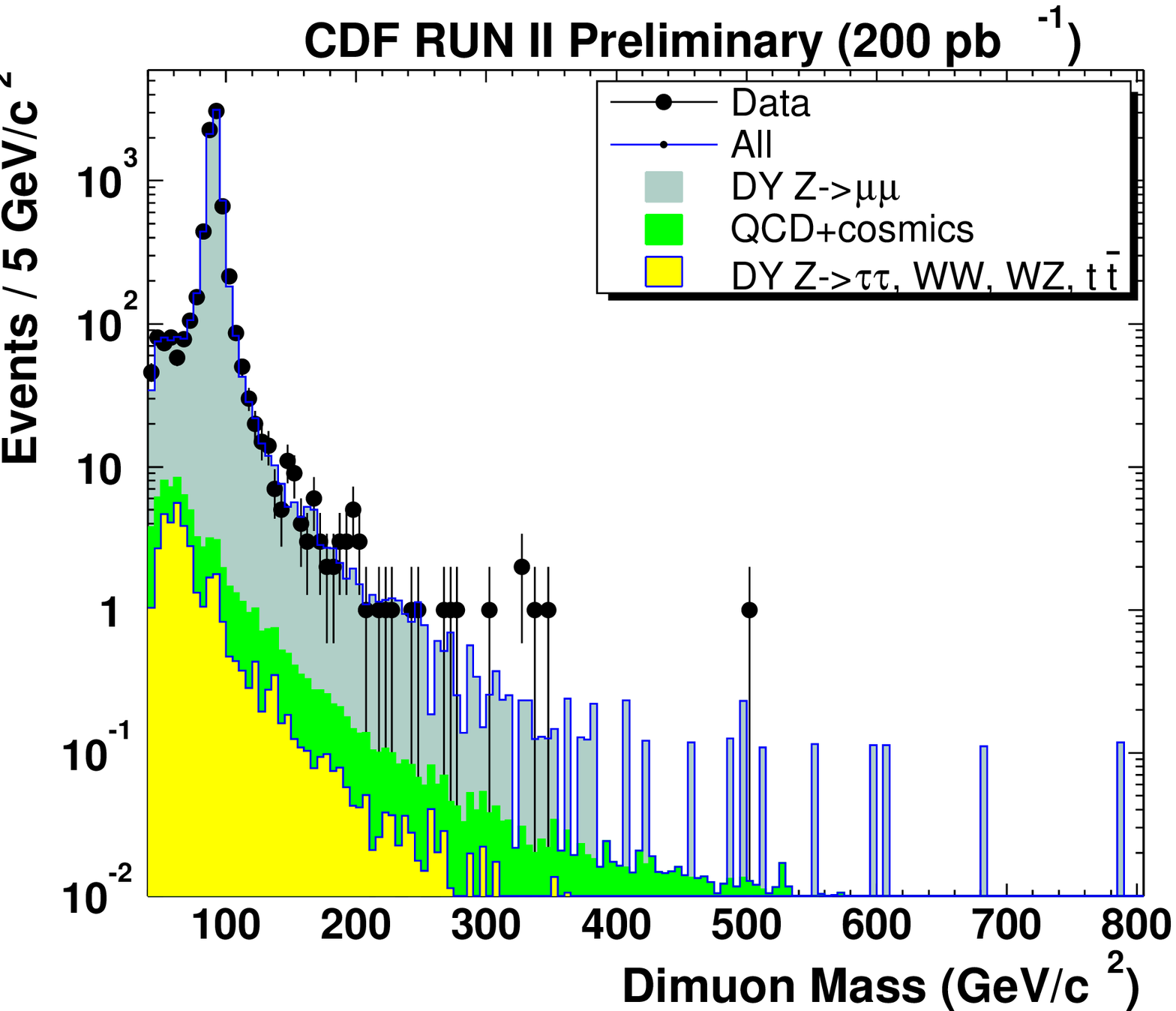}}
\caption{\label{cdf-uu} Di-muon invariant mass spectrum comparison
between data and estimated backgrounds at CDF Run II.
\hbox to258pt{}}
\end{figure}

\begin{figure}
\centerline{\epsfxsize 3.65 truein \epsfbox {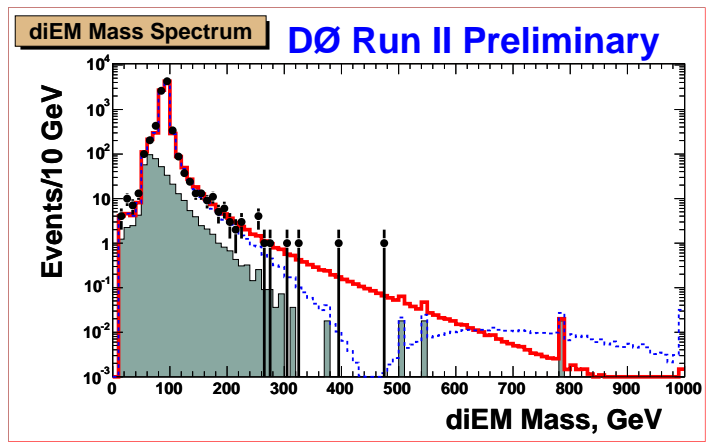}}
\caption{\label{d0-diem} Di-electron invariant mass spectrum comparison
between data and estimated backgrounds at D\O\ Run II with 200 pb$^{-1}$. The
dashed lines show the backgrounds plus the contribution from the TeV$^{-1}$ signal.
The deficit of expected events in the intermediate masses is due to negative
interference of the first KK $Z/\gamma$ mode with the Drell-Yan between
the $Z$ mass and $M_C$ (0.8 TeV).
\hbox to170pt{}}
\end{figure}

\begin{figure}
\centerline{\epsfxsize 3.65 truein \epsfbox {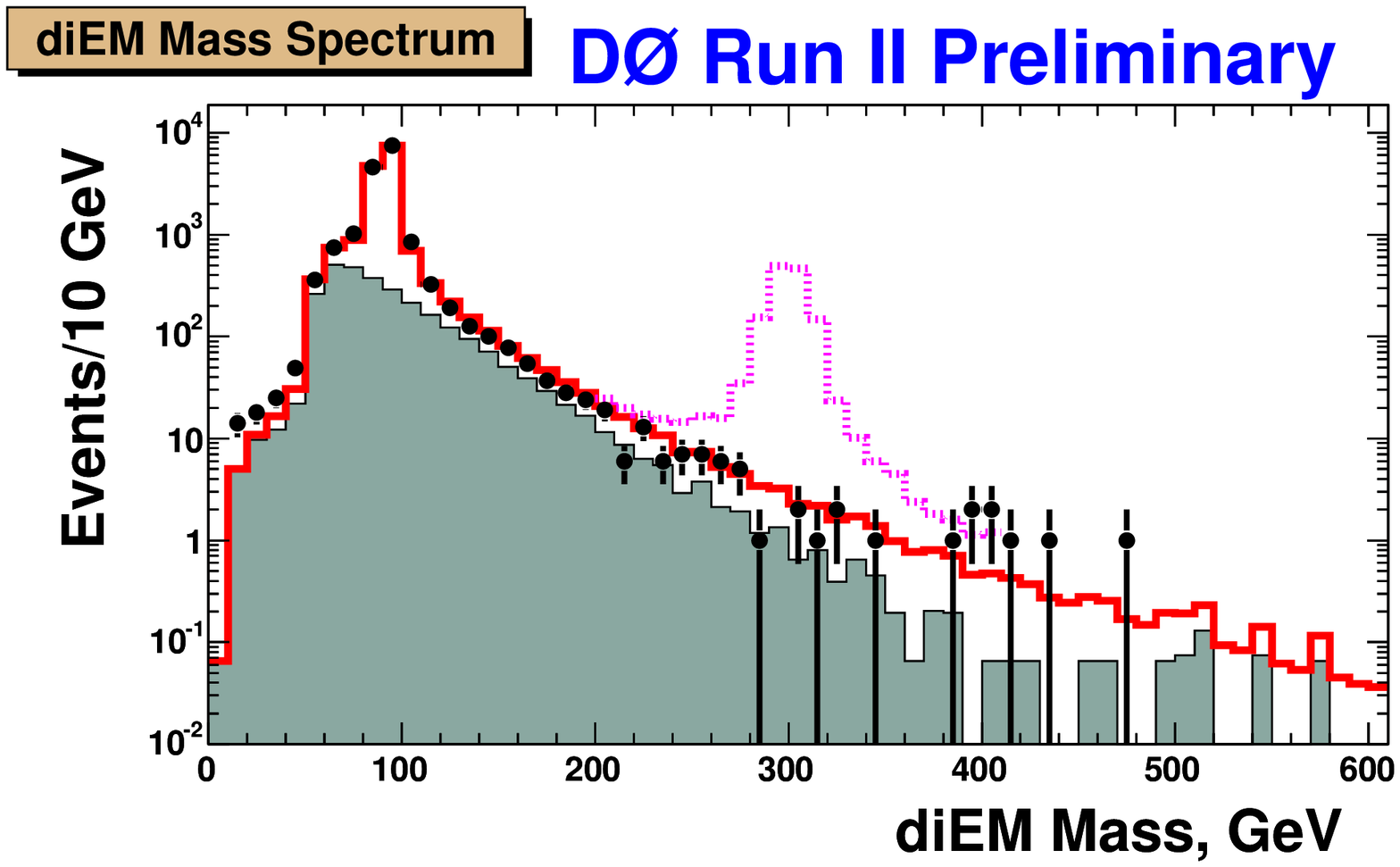}}
\caption{\label{d0-diem2} Di-EM (electron and photon candidates) invariant mass spectrum comparison
between data and estimated backgrounds at D\O\ Run II with 200 pb$^{-1}$. The shaded histogram indicates
misidentified jets as EM objects. A signal for a RS-graviton with mass 300 GeV/$c^2$ and arbitrary
cross section normalization is also shown.
\hbox to170pt{}}
\end{figure}

Representative spectra from CDF and D\O\ of di-lepton invariant mass spectra are
shown in figures \ref{cdf-ee}, \ref{cdf-uu}, \ref{d0-diem} and \ref{d0-diem2}. 
In the  D\O\ search both di-electrons and di-photons are considered simultaneously in 
the analysis and noted as ``diEM''. 

In the CDF di-lepton analyses the uncertainty on the total 
background estimate for $M_{\ell\ell}>300$ GeV/$c^2$ 
is 40\% for electrons, and 25\% for muons. 
Systematic uncertainties sources are  the luminosity, acceptance, 
energy scale and momentum resolution, selection efficiency, 
background statistics and normalization.

\begin{figure}
\centerline{\epsfxsize 3.45 truein \epsfbox {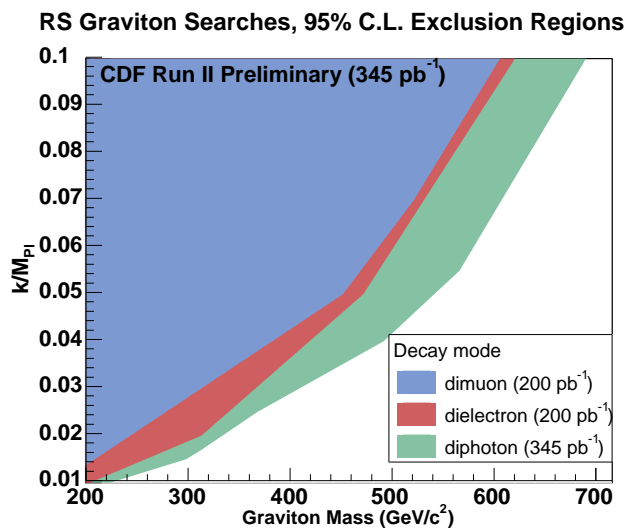}}
\caption{\label{cdf-rs-di-all} 95\% CL limits on the RS graviton mass-$k/M_{Planck}$ plane 
from the CDF search in three di-object channels as indicated. 
The shaded area to the left of the corresponding
curve is excluded.
\hbox to286pt{}}
\end{figure}

\begin{figure}
\centerline{\epsfxsize 3.35 truein \epsfbox {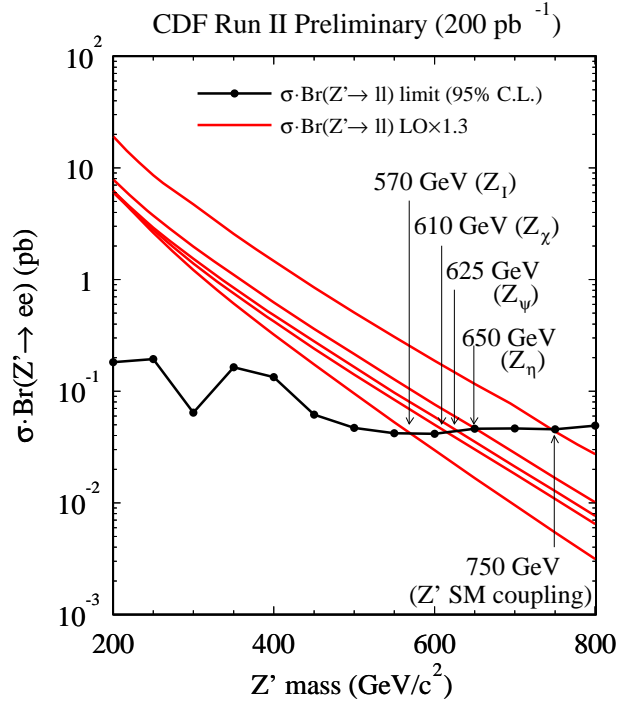}}
\vspace*{-0.3cm}
\caption{\label{cdf-zp} Upper bounds from 200 pb$^{-1}$ of CDF collected data
on the production cross section of a new $Z^{\prime}$ boson 
times branching ratio to decay into a di-electron
pair. Resulting bounds in several $E_6$ inspired $Z^{\prime}$ models are also
shown.
\hbox to170pt{}}
\end{figure}

\begin{figure}
\centerline{\epsfxsize 3.85 truein \epsfbox {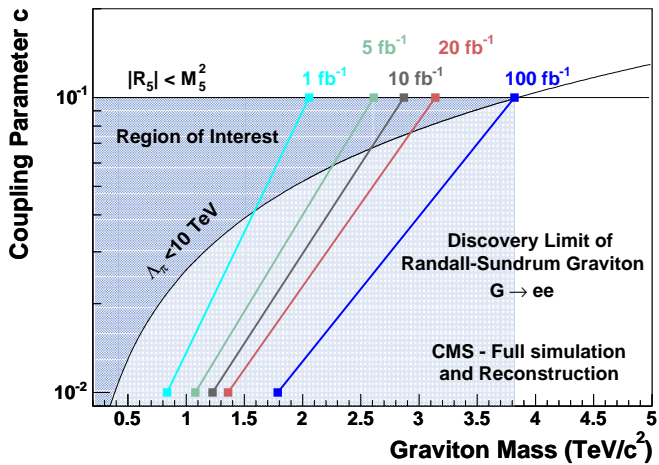}}
\caption{\label{caroline} CMS reach using the di-electron channel for
 RS gravitons, as a function of the ratio of
model parameters $\rm{c}=k/M_{Planck}$, and for integrated luminosities between 
1 and 100 fb$^{-1}$ \cite{caroline}.
\hbox to318pt{}}
\end{figure}

The null result in the high mass same flavor di-lepton/di-photon (and not shown here di-jet) 
channels at the Tevatron is  interpreted as 95\% CL limits in a variety of 
scenarios: ADD extra dimensions (estimated for 
Run II and the LHC in table \ref{di-add}; results from 200 pb$^{-1}$ 
of Run II are shown in table \ref{di-add2}), Randall-Sundrum gravitons \cite{lisa} (shown
in figure \ref{cdf-rs-di-all}),
a multitude of $Z^{\prime}$ models (shown in figure \ref{cdf-zp}) as well 
as technicolor particles and R-parity violating 
sneutrinos. It is interesting to note the reach improvement at the Tevatron in the 
case of the $Z^{\prime}$ search as a function of integrated luminosity: 
the 95\% CL limit on the mass was 505  GeV/$c^2$, 640 GeV/$c^2$
and $\sim$800 GeV/$c^2$  using 20 pb$^{-1}$, 90 pb$^{-1}$  and 200 pb$^{-1}$. The experiments
use either a fit of the di-lepton invariant mass  (CDF) or a mass window requirement and counting (D\O\ ).
A factor of 1.5 in mass reach is achieved with a factor of 10 in luminosity. 
At LHC (the examples are from CMS) less than 100 pb$^{-1}$, 
should be sufficient to discover $Z{^\prime}$ bosons of 1 TeV/$c^2$, 
a mass value which will likely be close to the final Tevatron reach.
For integrated luminosity of 100 fb$^{-1}$, the $Z^\prime$ discovery 
reach is in the range between 3.4 and 4.3 TeV (no systematics are considered in these estimates)
\cite{jason}. 
In the case of the di-electron final state analyzed in the context of RS gravitons, CMS
with an integrated luminosity of 100 fb$^{-1}$, CMS will cover the  region indicated
in figure \ref{caroline}.

\begin{table}
\medskip
\begin{center}
 \begin{tabular}{lccc}
 & \multicolumn{3}{c}{$M_S\;\;$ (TeV)}\\
 &  $n=2$ & $n=4$ & $n=6$    \\
\hline \hline
&\multicolumn{3}{c}{\underline{$p\bar{p}$, $\sqrt{s}=2$ TeV, 2 fb$^{-1}$}} \\
Di-Lepton & 1.9 & 1.6 & 1.3   \\
Di-Photon & 2.4 & 1.9 & 1.6   \\
Combined & 2.5 & 1.9 & 1.6 \\
\hline
& \multicolumn{3}{c}{\underline{LHC, $pp$, $\sqrt{s}=14$ TeV, 100 fb$^{-1}$}}  \\
Di-Lepton & 10 &  8.2 &  6.9  \\
Di-Photon & 12 &  9.5 &  8.0 \\
Combined & 13  & 9.9 & 8.3 \\
\end{tabular}
\caption{\label{di-add}Estimated sensitivity reach on the ultaviolet cuttoff $M_S$ at the Tevatron Run II and at the LHC \cite{tev-5}.}
\end{center}
\end{table}

\begin{table}
\begin{small}
\begin{tabular}{l@{}ccccccccc}
\hline
& GRW~\cite{GRW} & \multicolumn{6}{c}{HLZ~\cite{HLZ}} & \multicolumn{2}{c}{Hewett~\cite{Hewett}} \\
\hline
& & $n$=2 & $n$=3 & $n$=4 & $n$=5 & $n$=6 & $n$=7 & ~~$\lambda=1$ & $-1$\\
CDF Run II          & 1.06 &      & 1.32 & 1.11 & 1.00 & 0.93 & 0.88 & 0.99 & 0.96  \\
\hline
D\O\ Run II         & 1.36 & 1.56 & 1.61 & 1.36 & 1.23 & 1.14 & 1.08 & 1.22 & 1.10  \\
\hline
D\O\ Runs I+II & 1.43 & 1.67 & 1.70 & 1.43 & 1.29 & 1.20 & 1.14 & 1.28 & 1.14  \\
\hline
\end{tabular}
\caption{\label{di-add2}95\% CL lower limits on the ultraviolet cutoff $M_S$(in TeV) 
from the Tevatron Run II, 
within several phenomenological frameworks. NLO QCD effects have been accounted for
(signal and background) via a  $K$-factor of 1.3. \hbox to272pt{}}
\end{small}
\end{table}

\section{Mono-objects}
Both in Run I \cite{CDF-smaria-kev}, \cite{D0-monojet} and Run II \cite{greg-ecole} the Tevatron experiments 
use the missing energy plus a single jet as a probe
for Kaluza Klein gravitons in the ADD scenario via
the direct emission diagrams.
The on-shell production of
Kaluza-Klein gravitons produces a smooth missing energy 
distribution after convolution of the closely spaced KK
spectrum with the PDFs. This, coupled with the large
systematic uncertainties due to the jet energy scale
and the highly polluted with instrumental backgrounds 
missing energy triggers,
renders the channel challenging.

\begin{table}
\begin{tabular}{lccccccc}
\hline
  & $n=2$ & $n=3$ & $n=4$ & $n=5$ & $n=6$ & $n=7$ & $n=8$ \\
\hline
CDF\ mono-photons &       &       & 0.549 &       & 0.581 &       & 0.602 \\
\hline
D\O\ \ \hbox to3pt{}mono-jets  & 0.89  & 0.73  & 0.68  & 0.64  & 0.63  & 0.62 \\
\hline
CDF\ mono-jets   & 1.00  &       & 0.77  &       & 0.71 \\
\hline
\end{tabular}
\caption{\label{table:RunI} Individual 95\% CL lower limits on the fundamental Planck scale 
$M_D$ (in TeV) in the ADD model from the Run I data collected with the CDF and D\O\ experiments ($K$=1).
\hbox to330pt{}}
\end{table}

\begin{figure}
\centerline{\epsfxsize 3.55 truein \epsfbox {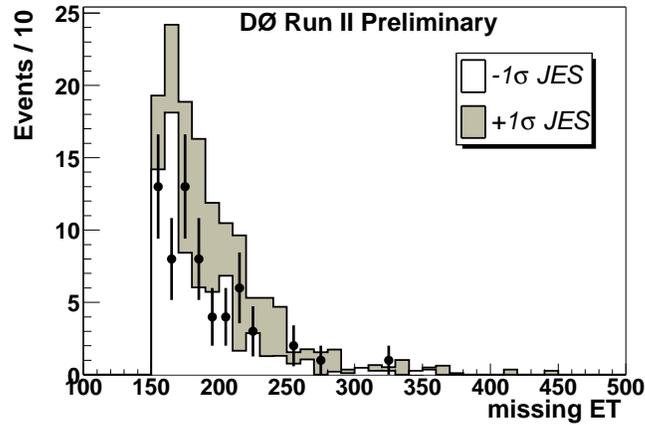}}
\vspace*{-0.3cm}
\caption{\label{d0-mono-met} Distribution of the missing $E_T$ for 
85 pb$^{-1}$ of Tevatron data at $\sqrt{s}=1.96$ TeV, 
collected with the D\O\ detector (points) and for non-QCD 
standard model background. The shaded band indicates the
effect of the jet energy scale uncertainties.\hbox to10pt{}
\hbox to170pt{}}
\end{figure}

\begin{figure}
\centerline{\epsfxsize 3.55 truein \epsfbox {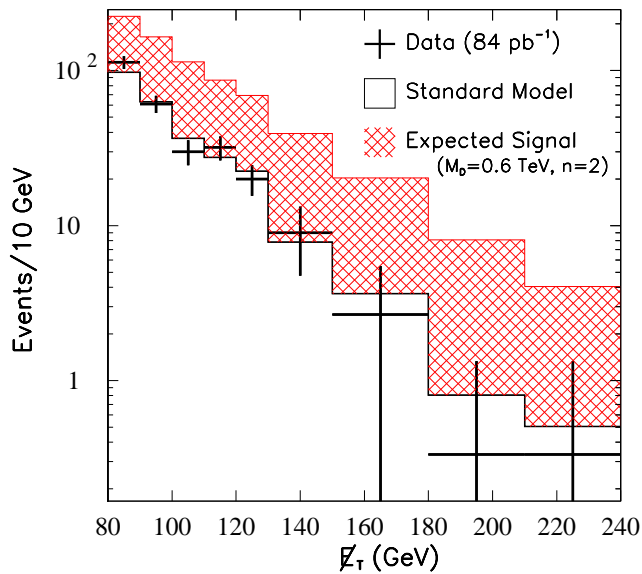}}
\vspace*{-0.8cm}
\caption{\label{cdf-mono} The predicted missing $E_T$ distribution from 
Standard Model  processes (histogram) and
the one from the expected graviton signal (for $n=2$, $M_{D}$ = 0.6~TeV, and
a $K$-factor of 1.0) added to the Standard Model background (hatched).
The signal appears as a smooth excess over the  background. The points are
the observed data at $\sqrt{s}=1.8$ TeV collected at CDF Run IB.
\hbox to303pt{}}
\end{figure}

\begin{figure}
\centerline{\epsfxsize 3.75 truein \epsfbox {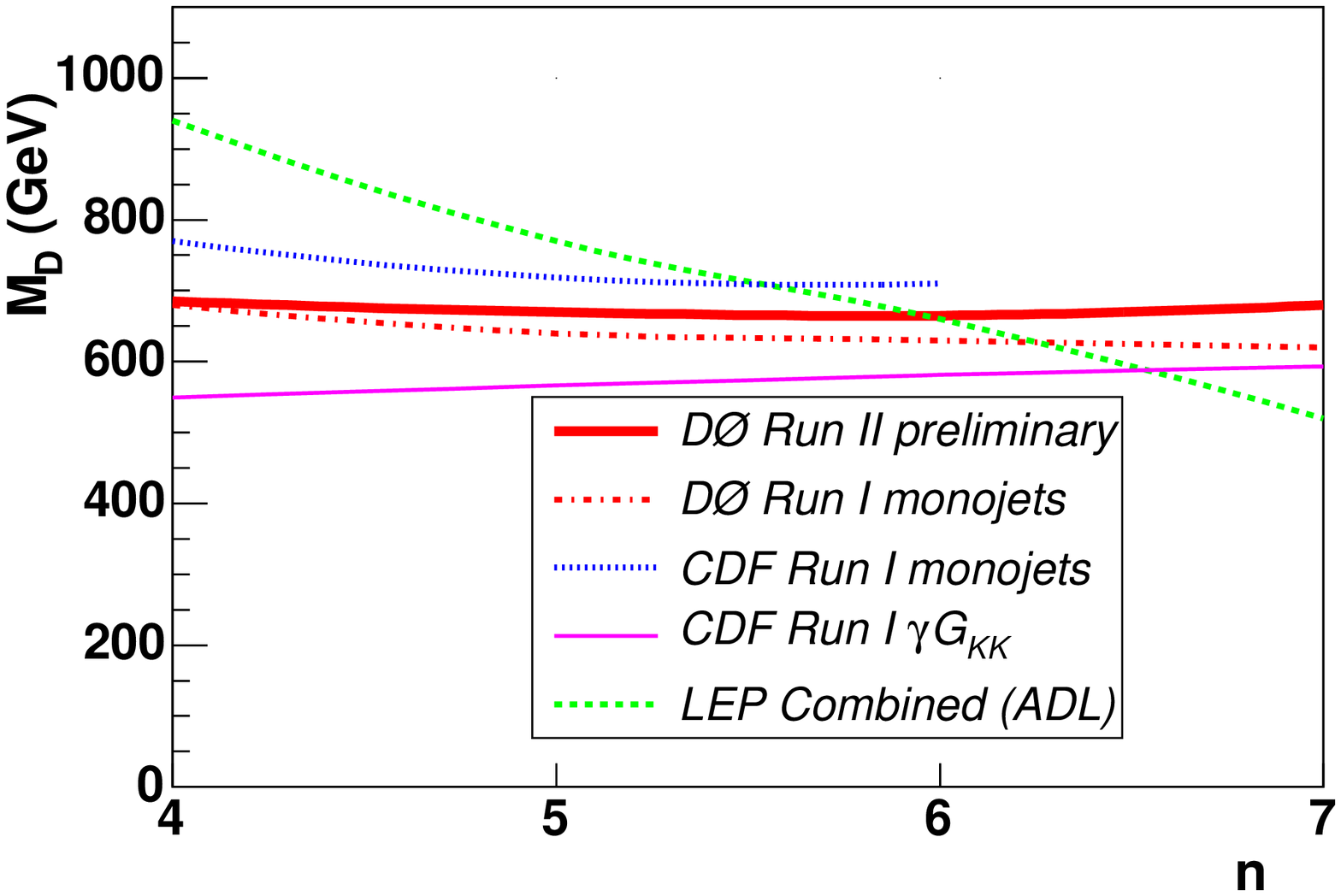}}
\vspace*{-0.5cm}
\caption{\label{monojet2} The limits on the effective Planck scale $M_D$ as a function
of the number of extra dimensions from: the Run II D0 monojet analysis (full line),
the CDF Run I (dotted line), D0 Run I (dashed-dotted line) and LEP combined (ADL) limit
(dashed line).
\hbox to42pt{}}
\end{figure}

The 95\% CL lower limits on the fundamental Planck scale $M_D$ (in TeV) in the ADD model from
85 pb$^{-1}$ of $p\bar{p}$ collisions at $\sqrt{s}=1.96$ TeV, collected by the D\O\ experiment in the 
monojet+missing energy channel and for $n=4,5,6,7$ extra dimensions are  0.68, 0.67, 0.66 and 0.68 TeV
correspondingly \cite{greg-ecole}. The missing $E_T$ distribution is shown in figure
\ref{d0-mono-met} \cite{D0-monojet}. The corresponding spectrum from 84 pb$^{-1}$ from Run I at CDF 
is shown in figure \ref{cdf-mono} \cite{CDF-smaria-kev} and 
the summary of all the results in the mono-jet and
mono-photon \cite{CDF-peter} analyses from the Tevatron and LEP is 
shown in figure \ref{monojet2} \cite{greg-ecole}.
Note that LEP is more sensitive for small number of extra dimensions and the Tevatron
takes over in sensitivity above 6 extra dimensions, with the jet channel being superior
to the photon one.

\section{Examples of Beyond SUSY searches at the LHC}

The LHC experiments in preparation for the upcoming physics
run have developed an extended program of exotic searches.
Here is an indicative signature- and model-based listing 
of Exotics (Beyond the Standard Model and SUSY physics) that is being 
investigated in LHC's ATLAS  (again circa July 2004):

\begin{itemize}
\item    Jets and Missing ET : 
 \begin{itemize}
        \item
         Signals of models with large extra dimensions in ATLAS
	 \item Graviscalars in ATLAS 
   \end{itemize}

\item   Narrow Graviton Resonances
\item   Virtual Graviton Exchange
\begin{itemize}
  \item      
           Di-photon, di-lepton, di-jet, 
         $t\bar t$ production from virtual graviton exchange
          
\end{itemize}

\item    Radion and other scalars
\begin{itemize}
\item
        Search for the Randall Sundrum radion using the ATLAS detector  
     \item    
        Graviscalar in ATLAS
        
\end{itemize}
\item    Gauge Excitations in TeV$^{-1}$ extra dimensions
\begin{itemize}
      \item 
                       KK excitation of the $W$ boson
      \item
                        KK excitations of gluons
\end{itemize}
\item     Black Holes
\begin{itemize}
        \item Black hole production and decay 
         \item Search for Gauss-Bonnet black holes
\end{itemize}
\item    Trans-Planckian Elastic Collisions

\item Singlet Neutrino
	\begin{itemize}
	\item          Hadronic tau decay of a heavy charged Higgs in models with singlet neutrino in large extra dimensions
\end{itemize}
\item    Dark energy
\begin{itemize}
\item        Dark Energy Signals and Cosmological Constant Signatures in ATLAS (in the contaxt of RS extra dimensions scenarios)

\end{itemize}
\item     Universal Extra Dimensions
\begin{itemize}
\item
        Di-jets in a scenario of Universal Extra Dimensions 
\end{itemize}
\end{itemize}


\section{Remarks}

An era of discovery is approaching with the onset of collisions at the Large Hadron Collider.
The searches for new phenomena in the currently running Tevatron are setting the
stringest limits on many models and with increasing luminosity the 
exploration of the TeV scale is well underway. The results (of which only a very
limited subset is presented here) as well as the problems faced and solved
at the  Tevatron experiments serve in many a case as guides to the  strong search and 
discovery program being developed at the LHC experiments \cite{LH}. 

\bigskip

{\small With many thanks to the organizers and exotic conveners of all the
collider collaborations, especially G. Polesselo, D. Denegri, L. Pape, G. Azuelos, J-F Grivaz and S. Lammel. Also to Stephan Lammel, Joe Lykken and
Albert De Roeck for a careful look at the manuscript and to  Claudia-Elizabeth  Wulz
for her insistence and patience.}


\bigskip

\begin{thebibliography}{99}
\bibitem{vienna}http://wwwhephy.oeaw.ac.at/phlhc04/index.html 
\bibitem{nature} The D\O\ Collaboration, Nature 429 (2004) 638-642
\bibitem{cdf-top} The CDF Collaboration, http://www-cdf.fnal.gov/physics/new/top/top.html
\bibitem{PL}P. Langacker, arXiv:hep-ph/0503068 (these proceedings).
\bibitem{JL-vienna}J. Lykken, arXiv:hep-ph/0503148 (these proceedings).
\bibitem{JS} J.L.~Hewett and M.~Spiropulu, Ann. Rev. Nucl. Part. Sci. {\bf 52}, 397 (2002).
\bibitem{SUSY} For a review see S.P. Martin in {\it Perpectives on Supersymmetry}, editor Kane, G.L. (1997)
[arXiv:hep-ph/9709356].
\bibitem{UED}
T.~Appelquist, H.~C.~Cheng and B.~A.~Dobrescu,
Phys.\ Rev.\ {\bf D64} (2001) 035002 [arXiv:hep-ph/0012100].
\bibitem{lisa} L. Randall and R. Sundrum, Phys. Rev. Lett. {\bf 83}, 3370 (1999);
L. Randall and R. Sundrum, Phys. Rev. Lett. {\bf 83}, 4690 (1999).
\bibitem{JR} For a review see J. Hewett, T. Rizzo, Phys. Rept {\bf 183}, 193 (1989).
\bibitem{split-savas} N. Arkani-Hamed, S. Dimopoulos, hep-th/0405159.
\bibitem{split-savas1} J.L.Hewett, B. Lillie, M. Masip, T. G. Rizzo, JHEP 0409:070,2004.
\bibitem{ADD}
N. Arkani-Hamed, S. Dimopoulos, G. Dvali, Phys. Lett. {\bf B429}, 263 (1998);
I. Antoniadis, N. Arkani-Hamed, S. Dimopoulos, and G. Dvali, Phys. Lett. {\bf B436}, 257 (1998);
N. Arkani-Hamed, S. Dimopoulos, G. Dvali, Phys. Rev. D {\bf 59}, 086004 (1999);
N. Arkani-Hamed, S. Dimopoulos, J. March-Russell, SLAC-PUB-7949, [arXiv:hep-th/9809124].
\bibitem{ignatios} I.~Antoniadis,
Phys.\ Lett.\ B {\bf 246}, 377 (1990); I.~Antoniadis, K.~Benakli, and M.~Quiros, Phys. Lett B {\bf 460}, 176 (1999).
\bibitem{tev-5}
K. Cheung and G. Landsberg, Phys. Rev. {\bf D65}, 076003 (2002).
\bibitem{greg-ecole} G. Landsberg, arXiv:hep-ex/0412028.
\bibitem{balazs}
C. Bal\'{a}zs and B. Laforge, Phys. Lett. {\bf B525}, 219 (2002).
\bibitem{jason} R.Cousins {\it et. al} CMS-CR-2004-050.
\bibitem{caroline} C. Collard {\it et. al} CMS NOTE-2004/024.
\bibitem{GRW}
G. Giudice, R. Rattazzi, and J. Wells, Nucl. Phys. {\bf B544}, 3 (1999),
[arXiv:hep-ph/9811291].

\bibitem{HLZ}
T. Han, J. Lykken, and R. Zhang, Phys. Rev. D {\bf 59}, 105006 (1999), [arXiv:hep-ph/9811350].

\bibitem{Hewett}
J. Hewett, Phys. Rev. Lett. {\bf 82}, 4765 (1999).

\bibitem{CDF-smaria-kev}
T.~Affolder {\it et al.\/} (CDF Collaboration), Phys. Rev. Lett. {\bf 92}, 121802 (2004).


\bibitem{D0-monojet}
V.~Abazov {\it et al.\/} (D\O\ Collaboration), Phys. Rev. Lett. {\bf 90}, 251802 (2003).

\bibitem{CDF-peter}
T.~Affolder {\it et al.\/} (CDF Collaboration), Phys. Rev. Lett. {\bf 89}, 281801 (2002).

\bibitem{LH}
B.~C.~Allanach {\it et al.}  [Beyond the Standard Model Working Group
                  Collaboration],
[arXiv:hep-ph/0402295].



\end{thebibliography}
\end{document}